# Temporal Psychovisual Modulation: a new paradigm of information display

Xiaolin Wu and Guangtao Zhai
Department of Electrical & Computer Engineering
McMaster University, Hamilton, Canada L8S 4K1

**Abstract** *We report on a new paradigm of information display that greatly extends the utility and versatility of current optoelectronic displays. The main innovation is to let a display of high refresh rate optically broadcast so-called atom frames, which are designed through non-negative matrix factorization to form bases for a class of images, and different viewers perceive self-intended images by using display-synchronized viewing devices and their own human visual systems to fuse appropriately weighted atom frames. This work is essentially a scheme of temporal psychovisual modulation in visible spectrum, using an optoelectronic modulator coupled with a biological demodulator.*

Information display technologies have played an indispensible role in the advancement of human civilization, from rock carving of prehistoric tribesmen to pervasive use of electronic displays in modern societies. The ever-presence, great variety and continuously increasing sophistication of electronic displays profoundly shape the ways we learn, work, communicate, entertain, travel, etc. and even how we socially behave. This article presents a new paradigm of information display that is conceived by making an ingenious interplay of critical flicker frequency of human vision, high refresh-rate digital display and optoelectronic viewing devices. The new display paradigm, called temporal psychovisual modulation (TPVM), differs fundamentally, in design principle, user experience and cost effectiveness, from the head-mounted display (HMD) technology (1,2), while offering the same functionalities and more.

TPVM exploits a well known phenomenon in psychophysics of vision: the human visual system (HSV) cannot resolve temporally rapidly changing optical signals beyond flicker fusion frequency (around 60 Hz for most viewers and under most conditions (3)). For the purposes of this research, the good news is that modern digital displays can run at much higher refresh rates, e.g. 120 Hz, 240Hz and beyond. For example, the new light modulators such as the deformable mirror devices and grating light valve devices can lead to very high refresh rate (up to 88k Hz) and spatial resolution for digital projectors; the main stream LCD/ LED technology also achieves 120Hz or 240Hz refresh rate, as demanded by emerging applications of 3D video playback. If the refresh rate of a digital display exceeds the limit of temporal resolution of HSV, then it is possible for HVS to "render" or perceive different images by fusing differently weighed consecutive frames (called *atom frames*) emitted from the same display. These psychovisual signal processing operations and the resulting net visual effects can be mathematically modeled as follows.

Let $f_d$ be the refresh rate of a digital display and $f_v$ the critical flicker frequency at which HVS just fails to discern discrete frames. For simplicity, assume $f_d = Mf_v$, where $M$ is a positive integer. Suppose that $\mathbf{y}_1, \mathbf{y}_2, \ldots, \mathbf{y}_K$ are the $K$ images to be formed by HSV. These $K$ target images constitute an $N \times K$ matrix $\mathbf{Y}$ with column $k$ being image $\mathbf{y}_K$ of $N$ pixels. Temporal psychovisual modulation is a problem of signal decomposition $\mathbf{Y} = \mathbf{XW}$ where $\mathbf{X}$ is an $N \times M$ matrix whose columns are $M$ atom frames $\mathbf{x}_1, \mathbf{x}_2, \ldots, \mathbf{x}_M$ of $N$ pixels to be cyclically displayed at frequency $f_d$, and $\mathbf{W}$ is an $M \times K$ modulation weighting matrix. In other words, image $\mathbf{y}_k, 1 \leq k \leq K$, is psychovisually formed as a linear combination of the $M$ atom frames. In the above problem formulation and in the sequel an assumption is made that the display is psychovisually linearized.

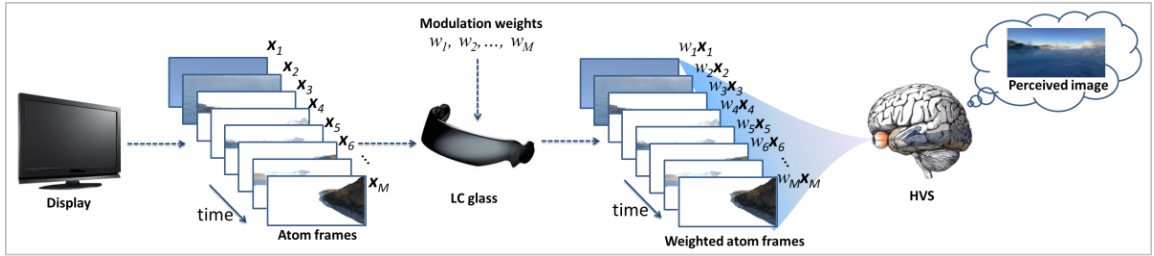

(A)

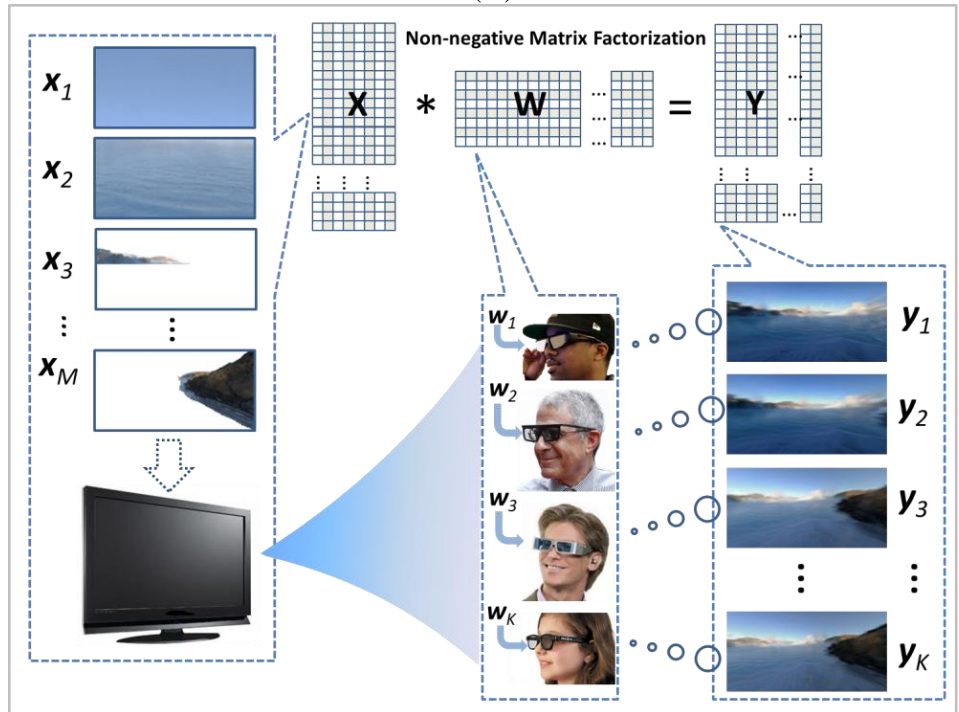

(B)

Fig. 1. Image formation via temporal psychovisual modulation in visible spectrum. (A) A high-speed display of refresh rate $f_d$ Hz emits $M = f_d / f_v$ atom frames, where $f_v$ is the critical flicker frequency. The light fields of these $M$ atom frames pass through and get amplitude-modulated by an active LC glass. The $M$ modulated atom frames are temporally fused by HVS and perceived as an image. (B) Concurrent multiple self-intended exhibitions via a common display medium, an optoelectronic-psychovisual process modeled by NMF. Different images $\mathbf{y}_1, \mathbf{y}_2, \ldots \mathbf{y}_K$ are perceived by different viewers whose LC glasses receive the same sequence of atom frames $\mathbf{x}_1, \mathbf{x}_2, \ldots \mathbf{x}_M$ (essentially basis functions) but are driven by different modulation weighting vectors $\mathbf{w}_1, \mathbf{w}_2, \ldots \mathbf{w}_K$.

TPVM can be implemented, as depicted in Fig.1, by a combination of a high-speed display emitting atom frames and display-synchronized active liquid crystal (LC) glasses that attenuate the light strengths of atom frames according to the modulation weights in $\mathbf{W}$. The defining and unique characteristic of the new TPVM display system is that the optoelectronic modulator is coupled with a biological demodulator, as HSV forms the image through temporal fusion of weighted light fields. Since the pixel values cannot be negative and active LC glasses cannot implement negative weights, the signal decomposition $\mathbf{Y} = \mathbf{XW}$ has to be a non-negative matrix factorization (NMF) (4), and moreover the weights in $W$ have to be in $[0,1]$. Algorithmically for the purpose of TPVM, we solve the following optimization problem

$$\min_{\mathbf{X},\mathbf{W}} \|\mathbf{Y} - \mathbf{XW}\|_F \quad subject\ to\ 0 \leq \mathbf{W} \leq 1, 0 \leq \mathbf{X} \leq 1$$

to find the $M$ atom frames that drive the high-speed display and the modulation matrix $W$ that drives (via wireless or wired links) the display-synchronized active LC glasses, with the objective of minimizing the errors in reconstructing the $K$ target images. The reconstruction errors can be made small for sufficiently large $M$ (i.e., high refresh frequency $f_d$), and if images $\mathbf{y}_1, \mathbf{y}_2, \ldots, \mathbf{y}_K$ are similar to one the other, such as in the case of 2D projections of the same 3D scene with respect to smoothly changing viewing angles.

In the TPVM design, a display of high refresh rate broadcasts atom frames, and different viewers can autonomously perceive self-directed images by using display-synchronized modulated viewing devices and their own HSV to fuse the atom frames, optoelectronically and psychovisually in cascade, as depicted by Fig. 1B. This process endows the TPVM display paradigm with the attractive functionality of concurrent multiple exhibitions on a same single screen, which greatly extends the usability of digital displays. When used in applications of virtual reality (VR) and augmented reality (AR) (1,2), TPVM can provide, in conjunction with motion sensors, multiple viewers with different real-time self-centered eyetracking perspectives; for example, a group of friends have a virtual walkthrough of a scenic spot. Comparing to current output devices of VR and AR (e.g., HMD) which exclusively serves a single user at a time, the new TPVM display system allows multiple users to interact with the same virtual world and at the same time enjoy each other's physical presence, offering a bodily experience of mutual participation that is lacking with HMD.

By delegating a bulk of imaging computations performed in HMD to HSV, the user-worn viewing apparatus of a TPVM-based VR/AR system becomes much simpler, lighter and far less expensive than HMD. Moreover, the TPVM display paradigm greatly reduces the computation power and video memory bandwidth required by real-time VR/AR applications, because a small number of atom frames, which can be precomputed for a particular 3D environment, are used to synthesize a range of different perspective views through appropriate attenuations of the active viewing devices and psychovisual processing of HSV. Indeed, on a frame for frame comparison the number of modulation weights to be communicated to viewing devices is negligible versus the number of pixels to be redrawn in the current practice.

In this work the NMF decomposition of the visual signal is stipulated by the hardware working mechanism of TPVM. Its necessity to our design aside, NMF has an interesting cognitive aspect. In the past decade, NMF has generated a great deal of interests and enthusiasm in research communities of machine learning, computer vision, and multivariate statistical analysis (5,6) as a useful modeling tool for image analysis and synthesis. NMF puts into computational practice the theories that visual cognition is based on a brain representation of a scene in terms of parts (7,8).

The parts-based brain representation of the whole is supported by some psychological and physiological evidence (9,10,11). Of significance to this research are the potential capability of NMF to represent semantically meaningful parts of a scene as bases and the one-to-one correspondence between such a basis and an atom frame in the proposed TPVM display system. Granted, atom frames are cyclically exhibited in very short duration (8.3ms for 120 Hz displays) each, but a flashing atom frame (an NMF basis) likely contains a meaningful part (see Fig. 1B) in a scene to which viewers are primed more often than not. This is certainly the case for applications of VR and AR, where viewers know what they are to experience, say a familiar cityscape or landscape. Under the above described circumstances, some cognition research results suggest that the viewer can respond to transient visual stimulus (12,13), lending support for the good perceptual quality achievable by the TPVM display paradigm.

Owing to its ability of generating different visuals on the same physical medium, the TPVM display system makes it possible to design and conduct a new type of scientific experiments in human cognition, psychology and physiology. At the present there exists no tool or instrument that can collect sample data on human judgment and behavior when the subjects are exposed to different visual stimuli but their other sensory (auditory, tactile, olfactory, and/or gustatory) inputs are the same and they are allowed personal interactions (verbal, gestural, etc.) in the same physical presence. The TPVM display technology fills this void, and it can facilitate studies on whether and how visual imagery correlates to other perceptions in a natural social atmosphere. For instance, many subjects can drink, smell, comment, and then evaluate/identify the same wine in a party setting, while watching from a common big screen a video that they believe to be the same but made different to different individuals by TPVM.

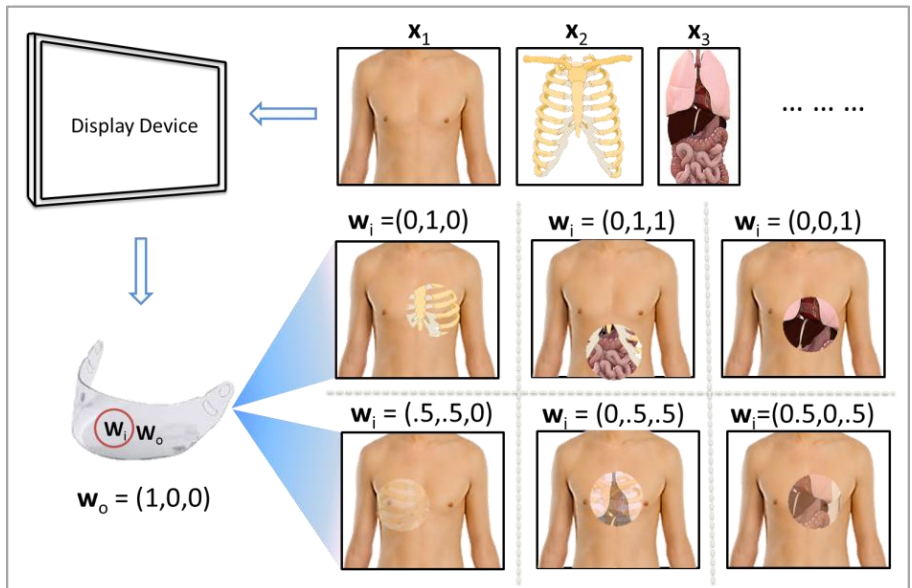

Fig. 2. Visualization effects achieved by altering modulation weights on the surface of active LC glass. Different users can choose different self-directed views simultaneously by moving their heads (or lines of vision), and different 3D regions of interest by adjusting modulation weights.

Some new VR/AR user experiences can be created, which are otherwise impossible with existing display technologies, by varying temporal modulation patterns of the synchronized active viewing devices in pixel locations. Fig. 2 shows such an example for 3D medical visualization.

Multiple layers of registered human anatomy images are mapped to atom frames and cyclically exhibited at high speed to doctors and students, each of whom sees the display through his/her own active LC visor. The surface of the LC visor is partitioned into inner and outer areas, which are respectively assigned modulation weighting vectors $\mathbf{w}_i = (w_{i1}, w_{i2}, \ldots, w_{iK}) \in [0,1]^K$ and $\mathbf{w}_o = (w_{o1}, w_{o2}, \ldots, w_{oK}) \in [0,1]^K$. By choosing different $\mathbf{w}_i$ and $\mathbf{w}_o$, a viewer can experience different desired see-through visual effects of the human body. For instance, by setting $\mathbf{w}_i = (0,0,1)$ and $\mathbf{w}_o = (1,0,0)$, a user sees a selected internal organ in alignment with the exterior of the human body. It is also straightforward to perform Alpha blending of anatomy layers by adjusting the modulation weights of the display-synchronized LC visor. Similarly, TPVM can be applied to visualize a raw 3D volume data set. Slices of the 3D volume, $\mathbf{x}_1, \mathbf{x}_2, \ldots, \mathbf{x}_M$, are cyclically displayed at high speed. The display-synchronized LC visor employs a concentrically varying modulation pattern, with which TPVM offers a funnel-shaped "dig-in" view into the 3D volume (mosaic of depth layers). The flexibility and effectiveness of the proposed TPVM medical visualization system are demonstrated in the online supporting material: www.ece.mcmaster.ca/~xwu/Demo4Science/TPVMdemo1.avi.

Although the above described 3D visualization tasks can be performed by software on a conventional display for a single user, it is not possible for multiple users, say a group of physicians, each of whom wants to autonomously examine the 3D data set while consulting one the other and reading the same screen. With the TPVM display system, each user can immerse into a 3D scene at a location of his choice simply by changing the line of vision and adjusting the weights of different depth layers, instead of being forced to have a particular view of someone else's, as shown in Fig. 2.

In addition, TPVM offers an intriguing perceptual mechanism of "seeing is not believing". To users wearing no modulation viewing devices the high-speed display exhibits the image $\mathbf{y}_0 = \mathbf{x}_1 + \mathbf{x}_2 + \ldots \mathbf{x}_M$ (all atom frames are fused unattenuated). This view rendered by HVS of all atom frames displayed in rapid succession and of equal weight is called *normal view*, which is our daily visual experience without tampering the lights from the display. Contrarily, other views perceived with unequal weighting of the atom frames are called *shale views*. The normal view $\mathbf{y}_0$ can be made drastically different from a shale view $\mathbf{y}_k$ perceived through modulated viewing devices. Such a bifurcation in psychovisual perception caused by whether to use a modulated viewing device lends TPVM to a diverse range of applications, including information security and privacy, consumer electronics, digital gaming, steganography and etc. These applications fall into two classes, categorized by the function of the normal view. In the first class, the normal view plays a role similar to covertext in steganography. Since a shale view is camouflaged by the normal view and made unrecognizable to unaided eyes, it can carry secrete messages that are only revealed through an encrypted modulated viewing device. Thus TPVM allows a user to read confidential information displayed on a personal device in public areas with no worry of being eavesread (Fig. 3A). This optoelectronic security solution offers much stronger protection than the privacy filter that can only restrict the viewing angles. If implemented on a touch screen of high refresh rate, TPVM can realize an invisible keypad on which only the user with matched modulation glasses can read and operate the soft buttons. A 2010 study found that 68% of users did not feel secure when typing passwords using a conventional keypad in public. The PTVM-based invisible keypad offers a perfect solution to the problem, and should find its way into daily life, installed on ATMs, store check-out counters, credit/debit card machines, combination locks and etc.

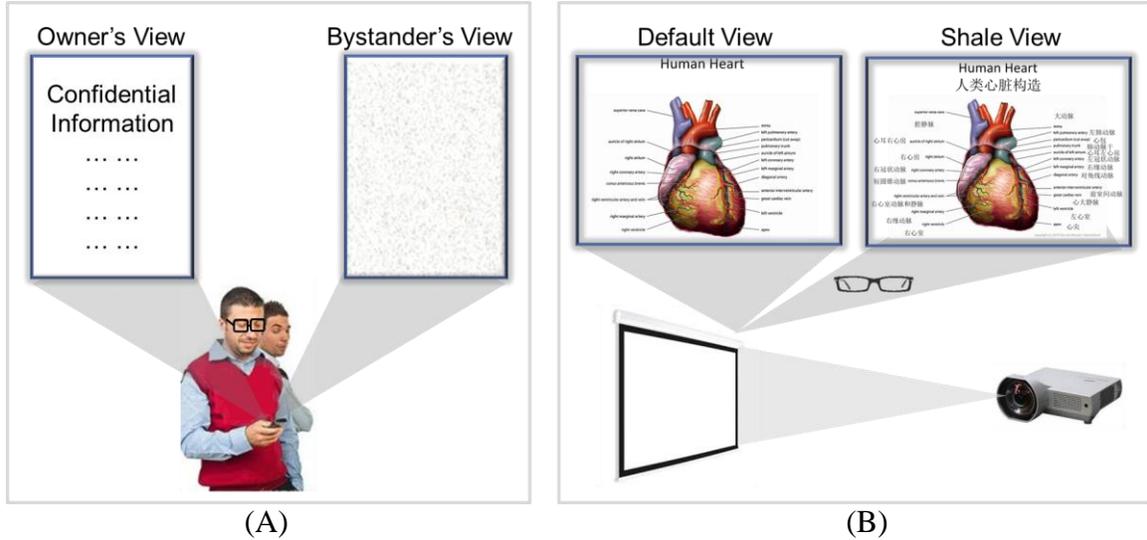

(A)  (B)

Fig. 3. Normal view vs. shale view. (A) Shale view contains confidential information that is visible only to the viewer using encrypted LC glasses, while the normal view is a cover image or decoy. (B) Shale view is an annotated normal view for viewers of special needs, while the normal view is the default image for unaided eyes.

Perceptual bifurcation for security applications can be achieved as the following special case of NMF, assuming $M = 2$ (the lowest possible)

$$(\mathbf{y}_0, \mathbf{y}_1) = (\mathbf{x}_1, \mathbf{x}_2) \begin{pmatrix} 1 & w_1 \\ 1 & w_2 \end{pmatrix}$$

with the secrete image $\mathbf{y}_1$ given, and the design variables being cover image $\mathbf{y}_0$, atom frames $\mathbf{x}_1, \mathbf{x}_2$, and the modulation weights of the viewing devices. Finding the optimal solution of NMF is NP-hard (14), but a good and real-time heuristic solution can be worked out. For the best legibility of the secrete message read through a modulated viewing device, we let $\mathbf{x}_1 = \mathbf{y}_1$, i.e., $w_1 = 1, w_2 = 0$. Then the remaining task is to choose atom frame $\mathbf{x}_2$ to best camouflage $\mathbf{y}_1$ by the resulting cover image $\mathbf{y}_0 = \mathbf{x}_1 + \mathbf{x}_2$. For example, we can simply make the normal view $\mathbf{y}_0$ appear a random noise image $\mathbf{n}$ by setting $\mathbf{x}_2 = \mathbf{n} - \mathbf{x}_1$. In practice, however, proper constraints should be in place so that the resulting pixel values in atom frame $\mathbf{x}_2$ always fall into the dynamical range of the display.

In the second class of applications, the normal view is not meant to conceal as in security applications but rather to be a default view for majority or casual viewers, whereas the shale view is for some user(s) of special needs who has to share the display with others. For example, often a public speaker (e.g., meeting presenters, teachers, entertainers, etc.) desires to follow but without appearing to read notes during his/her slides presentation. TPVM allows private notes to be projected onto the screen that are transparent to the audience but visible to the speaker only. In this case, the normal (default) view is the visual intended for the audience, whereas the shale view, which is an annotated normal view, can be seen by the presenter via a modulated viewing device. Similarly, TPVM can facilitate multilingual presentations. In the default view, the contents are displayed in an international language, say English; in a shale view, the contents of are displayed in a less common language, say Korean (Fig. 3B). Demonstrations of these applications of TPVM can be found in the online supporting material:
www.ece.mcmaster.ca/~xwu/Demo4Science/TPVMdemo2.avi.